\documentclass[preprint]{aastex63}
\usepackage{enumitem}

\newcommand\msun{~M$_{\odot}$}

\received{}
\revised{}
\accepted{}

\submitjournal{ApJ}

\shorttitle{Planetary nebulae and stellar abundances}
\shortauthors{Stanghellini, et al.}

\begin{document}

\title{Abundances of planetary nebulae and evolved stars: Iron and sulfur depletion, and carbon and nitrogen enrichment, in low- and intermediate-mass stellar populations in the Milky Way.}

\correspondingauthor{Letizia Stanghellini}
\email{letizia.stanghellini@noirlab.edu}

\author[0000-0003-4047-0309]{Letizia Stanghellini}
\affiliation{NSF's NOIRLab,
950 N. Cherry Ave., Tucson, AZ 85719, USA}

\author[0000-0002-0134-2024]{Verne V. Smith}
\affiliation{NSF's NOIRLab, 950 N. Cherry Ave., Tucson, AZ 85719, USA}
\affiliation{Institut d'Astrophysique de Paris, CNRS and Sorbonne Université, 75014 Paris, France}

\author[0000-0001-6476-0576]{Katia Cunha}
\affiliation{Steward Observatory, University of Arizona, Tucson, AZ 85721, USA}
\affiliation{Institut d'Astrophysique de Paris, CNRS and Sorbonne Université, 75014 Paris, France}
\affiliation{Observat\'orio Nacional MCTI, S\~ao Crist\'ov\~ao, Rio de Janeiro, Brazil}

\author[0000-0002-4591-6253]{Nikos Prantzos}
\affiliation{Institut d'Astrophysique de Paris, CNRS and Sorbonne Université, 75014 Paris, France}

\begin{abstract}
We explore the elemental abundances in Galactic planetary nebulae (PNe) compared with those of their stellar progenitors (Red Giant Branch and Asymptotic Giant Branch, RGB and AGB, stars), to explore and quantify the expected -- i.e., due to AGB evolution or condensation onto grains --- differences. We gleaned the current literature for the nebular abundances while we used the APOGEE DR~17 survey data for the stellar sample. We examined the elements in common between the nebular and stellar samples, namely, C, N, O, Fe, and S. 
We confirm that iron in PNe is mostly entrapped in grains, with an average depletion  $<$D[Fe/H]$>$=1.741$\pm$0.486 dex, and we disclose a weak correlation between iron depletion and the [O/H] abundance, D[Fe/H]$=(6.6003\pm2.443)\times{\rm [O/H]} +(1.972\pm0.199)$. Sulfur may also be mildly depleted in PNe, with $<$D[S/H]$>=0.179\pm0.291$ dex. 
We also found an indication of nitrogen enrichment for PNe $<$E[N/H]$>$=0.393$\pm$0.421 dex, with maximum enrichment (0.980$\pm$0.243) occurring for the PNe whose progenitors have gone through the HBB. The carbon enrichment is $<$E[C/H]$>$=0.332$\pm$0.460 dex when measured for the general PN populations. Our results will be relevant for future Galactic and extragalactic studies comparing nebular and stellar samples. 

\end{abstract}

\section{Introduction}

Planetary nebulae (PNe) are the post-Asymptotic Giant Branch (post-AGB) ejecta of evolved low- and intermediate-mass stars (LIMS, with initial mass in the $\approx$ 0.85-8\msun) range. 
The population of $\sim$2500  Galactic PNe (i.~e., the {\it true PNe} in the \citet{HASH} catalog) is distributed predominantly in the Galactic plane, with a limited bulge population and just a few PNe in the halo \citep{KH22,SH18}. The emission spectra of Galactic PNe allow the direct abundance analysis of several $\alpha$-processed elements (hereafter, $\alpha$-elements), such as, O, Ne, Ar, S, in about 20$\%$ of the known Galactic PNe  \citep[e.~g.][]{SH18}. Since these are distributed broadly at different Galactic radii, it is possible to determine a radial metallicity gradient from them \citep[e.~g.][]{BS23}, and can probe the chemical evolution of galaxies with look-back times from 0 to $\sim$10 Gyr, depending on the initial progenitor mass of the PNe \citep{Gibson2013,Magrini16,2019arXiv190504096S,2020MNRAS.492..821C}. 
Gas-phase abundances of $\alpha$-elements measured in PNe should, in principle, be the same as in their original stellar progenitors since the evolution of these elements is minimal in LIMS \citep[e.~g.][]{2017MNRAS.471.4648V}. 

Furthermore, PNe abundances of those elements (e.~g., C and N) that change as a star evolves along the Red Giant Branch (RGB) and on the AGB, ultimately probe stellar evolution in the PN progenitor mass range. The comparison between C and N abundances in PNe with those abundances in the progenitor stellar population probes stellar evolution since the PNe abundance of such elements are expected to differ from those of the progenitor stars due to, for example, dredge-up (DU) activity and hot-bottom burning (HBB) during the AGB phase \citep[see][]{Ventura15}.
The final C and N abundances depend strongly on the progenitor mass and (but less so) on
initial metallicity. The comparison of C and N in PNe and their progenitors lead to a better understanding of the evolutionary link between the two populations. Ideally, we can segregate the PNe according to their mass (i.e., age) and study them along with an underlying population of progenitor stars having similar ages.

A direct abundance comparison between Galactic PNe and their progenitor stellar population has been rarely, and not recently, done. The seminal paper by \citet{SL1990} analyzed CNO isotopes and mixing in a sample of 32 Red Giants (RGs)\footnote{Hereafter, the term ``Red Giants" (RGs) includes both RGB and AGB stars}, discussing their N and O abundance results in comparison with those obtained for a sample of nearly 100 PNe \citep{1979ApJ...228..163K,K1980,AC1983,1987ApJS...65..405A}. 
They found a reasonably good match for the abundances in the two populations, with the N and O abundance range in the RGs being roughly comparable to that of the PN sample. 

There is now a larger number of PN abundance measurements available in the literature to be considered. In addition, the last data release of the SDSS APOGEE survey DR~17 (Holtzman et al., in preparation) contains chemical abundances for several elements in hundreds of thousands of stars, offering a wealth of abundance results in many stellar types and a great opportunity to study abundance patterns in PNe and RGs, and to compare the results for the two samples seamlessly. With these data sets in hand, we will compare and discuss the abundances of C, N, O, S, and Fe in RGs and PNe. However, such comparisons must recognize that certain elements might condense into solid phases (with the solid phase assumed to be small ``dust'' particles), and these elements may suffer various amounts of depletion from their gas phase abundances. We will evaluate if there is a deficit between the abundances of PNe and those of the progenitor stars, particularly for those elements that do not vary during the stellar evolution of LIMS. These deficits in the elemental abundances occur because in PNe we can only measure gas-phase abundances, while some elements may be aggregating into dust grains. Most importantly, in this study elemental depletion can be measured in a more meaningful way, since it can be measured relative to the underlying stellar population rather than just relative to the solar abundances, or, simply compared with nebular abundances in H~II regions, which also suffer from depletion onto grains.

In Section 2, we present the selection criteria of the nebular and stellar samples. In Section 3 we discuss  the oxygen abundances, which are used as proxies for the metallicity indicators both in the PNe and stars. Section 4 focuses on the iron and sulfur depletion in PNe compared to the underlying stellar population. In Section 5, we discuss carbon and nitrogen evolution and enrichment in LIMs. Finally, Section 6 presents the discussion and concluding remarks.

\section{Data Samples} \label{sec:samples}

\subsection{Nebular Sample}\label{sec:pn_sample}

The comparison of the abundances of Galactic PNe and their progenitors stars starts with the selection of the nebular sample. Our initial PN catalog is the Hong Kong/AAO/Strasbourg H-alpha planetary nebula database, or HASH \citep{HASH}, where we extracted information for all (2556) {\it true} PNe. We then select PNe within the Galactic disk, limiting Galactocentric distances to $3<R_{\rm G}<35$ kpc and altitudes on the Galactic plane to $-2<z<2$ kpc, using to calculate these spatial parameters  the Galactic coordinates of the PNe, the solar distance to the Galactic center ($R_{\odot}=8.125$ kpc, \citet{2018A&A...615L..15G}), and the heliocentric statistical distances to the PNe, calibrated with Gaia DR3 parallaxes of their central stars (CSs) \citep{BS23}. However, the results in this paper would not change if we use instead the distance scale based on Gaia DR2 parallaxes \citep{Stanghellini2020} since the two sets of distances are comparable within the uncertainties. 

With these unique PNe in hand, we gleaned the literature for the abundances of nitrogen, carbon, oxygen, sulfur, and iron. We used only direct abundances, based on plasma diagnostics and emission line intensities, and most of them use the ionization correction factors (ICFs) to derive elemental abundances from ionic abundances. 
We select the reliable elemental abundances measurements from literature references, as described in \citet{SH18}, and added to the references therein
the data sets by \citet{TBL03,MHB19,McNabb2016,DIRP15,ADIGR20}. In papers where abundances are given both for the whole nebula and for different parts of the PNe, we use only those measurements that include the whole PNe, so the abundances from different references are readily comparable. We do not include in this study abundance entries with spectral data only covering part of the PN.

In Table~\ref{table:PN_abref} we list the references for PN abundances for the elements discussed in this paper.

\begin{table}
\caption{Nebular Abundance References}          
\label{table:PN_abref}     
\scriptsize
\begin{tabular}{ll}
\hline
{Element}& {References} \\
\hline

[O/H]&  KB94, KH01, MHW02, KHM03, TBL03, CUM04, PMS04, SGC06, PBS10, GHG14, DIRP15, DKS15, MHB19  \\

[N/H]&  KB94, KH01, MHW02, KHM03, TBL03, CUM04, PMS04, SGC06, PBS10, GHG14, DIRP15, DKS15, MHB19 \\

[S/H]& KB94, TBL03, PBS10, GHG14, MHB19  \\

[C/H]& KB94, TBL03, PBS10, GHG14, MHB19, SBS21\\

[Fe/H]& DIR1\\

\hline
\end{tabular}
\tablerefs{\citep[][KB94]{KB94}; \citep[KH01]{KH01}; \citep[MHW02]{MHW02}; \citep[KHM03]{KHM03}; \citep[TBL03]{TBL03}; \citep[CUM04]{CUM04}; \citep[PMS04]{PMS04}; \citep[SGC06]{SGC06}; \citep[PBS10]{PBS10}; \citep[DIR14]{DIR14}; \citep[GHG14]{GHG14}; \citep[DIRP15]{DIRP15}; \citep[DKS15]{DKS15}; \citep[MHB19]{MHB19}; \citep[SBS21]{SBS21}.}
\end{table}

Different authors may use different ICF schemes to supply information for missing emission lines that are expected in the nebular spectra. We produced a curated catalog of final abundances where we recalculated the ICFs uniformly, as described in \citet{SH18}. We use in this paper the curated C, N, and O abundances published in \citet{BS23}. In Table~\ref{table:abundances} we add the sulfur and iron abundances, and convert all entries 
in the [X/H] format, where we used the solar abundances from \citet{GAS2007}, the same used to scale DR~17 abundances. Uncertainties in the abundances are typically not given in the original references. We assume 20$\%$ uncertainty for all elemental abundances, except for sulfur where we assume 25$\%$ uncertainty, all based on the spread of abundances by different Authors when multiple references with errors are available.

\begin{table*}
\caption{Elemental Abundances of Galactic PNe}

\label{table:abundances} 
\begin{tabular}{lrrrrr}
\hline
PN~G &  [C/H]&  [N/H]& [O/H]& [S/H]& [Fe/H]\\
\hline
\hline
000.3+12.2 &      \dots& -0.100 &-0.160     & \dots&\dots\\
001.6-01.3  &     \dots& 0.670 & -0.390       & \dots&\dots\\
002.0-13.4 & \dots & 4.000 & 0.000 & \dots & \dots \\
\hline
\end{tabular}
\tablecomments{The Table is available online.}
\end{table*}

There are 694 {\it True} PNe in the HASH catalog which have statistical distance determination, populating the {\it initial PN sample} (see Table~\ref{Samples}). 
In the initial PN sample, 258 PNe have at least one [O/H] measurement in the literature. We call this the "[O/H] PN sample".  Similarly, we build the other [X/H] PN samples from the [O/H] PN sample by selected PNe that have at least one measurement of the [X/H] elemental abundance in the literature, with X=Fe, S, N, C. Given the procedure in building the samples, all PNe in the [X/H] samples have measured spatial ($R_{\rm G}$ and $z$) and metallicity ([O/H]) parameters. All PN samples used in this paper are described in Table~\ref{KS}.
  
\begin{table}
\caption{Nebular and Stellar samples}          
\label{Samples}     
\scriptsize
\begin{tabular}{lllllll}
\hline
Original Catalog& Sample& N& log(g)& $T_{\rm eff}$ & $R_{\rm G}$&  $|z|$\\
\hline
HASH \tablenotemark{a}& initial PN sample& 694& $\dots$& $\dots$& 3--35& $<$2 \\
APOGEE DR~17 \tablenotemark{b}& initial RG sample& 138,981& -0.5--2.5& 3500--5000& 3--35& $<$2\\ 
\hline
\end{tabular}
\tablenotetext{a}{Selecting only {\it true} PNe, with [O/H] measurements}
\tablenotetext{b}{Excluding {\it badbit} and [Fe/H]$>$-6 entries. Selecting S/N$>100$} 
\end{table}

\subsection{Stellar Sample} \label{sec:star_sample}

To select a sample of stars representing the PN progenitors we started with the APOGEE survey DR~17 \citep{2021AJ....162..302B}, a spectroscopical survey that includes accurate abundance determinations. We use the calibrated abundances directly from DR~17. Following the recommendations of the DR~17 tutorials (see https://www.sdss4.org/dr17/irspec/apogee-tutorials/), we ingested the complete stellar spectral data, eliminating the {\it badbit} entries; furthermore, we eliminate all entries that do not satisfy to the requisites $T_{\rm eff}>$0., log$(g)>-10$, [Fe/H]$>-6$, and S/N$>100$, i.e., we include in the selected sample only stars that satisfy the request that the stellar spectrum has high S/N, enough to produce meaningful values for the temperature, gravity, and abundances, to assure the best quality data selection.

Based on Galactocentric distances ($R_{\rm G}$) and distances from the Galactic mid-plane ($z$) of the stars from the DR~17 Value Added Catalogue ASTRONN \citep{LB2019}, we selected stars that are in the Galactic disk by imposing the constraints $3<R_{\rm G}<35$ kpc and $-2<z<2$ kpc. 
We restricted the stellar sample to cover RGs by constraining the stellar parameters as follows: $3500<T_{\rm eff}<5000$ K, and $-0.5<{\rm log}(g)<2.5$.  These intervals result from restricting the sample parameters to include primarily the low- to intermediate-mass red giant (both RGB and AGB) progenitors of PNe, where M$<$8~\msun, as more massive stars evolve through the supergiant phase to become supernovae of Type II \citep[][]{KL2014}.  The range in T$_{\rm eff}$ includes most of the RGB, RC, and AGB stars, while the lower limit in log(g) excludes the more luminous, massive supergiants.  Since there is a luminosity upper limit to AGB stars, due to their upper core-mass limit of $\sim$1.4\msun, of L/L$_{\odot}\sim 5\times10^4$ to 6$\times10^4$, depending on the metallicity, coupled with the maximum mass \citep{2004MNRAS.350..407I}, there is a lower limit to the expected surface gravity of log(g)$>$-0.4, and we apply a minimum of -0.5 for log(g) to delineate approximately the surface gravity limit for AGB stars.

This selection resulted in a sample with $\sim$140,000 stars which we will refer to as the {\it initial RG sample}. The selection criteria for this sample are given in Table~\ref{Samples}.

\begin{figure}
   \centering
   \includegraphics[width=18cm]{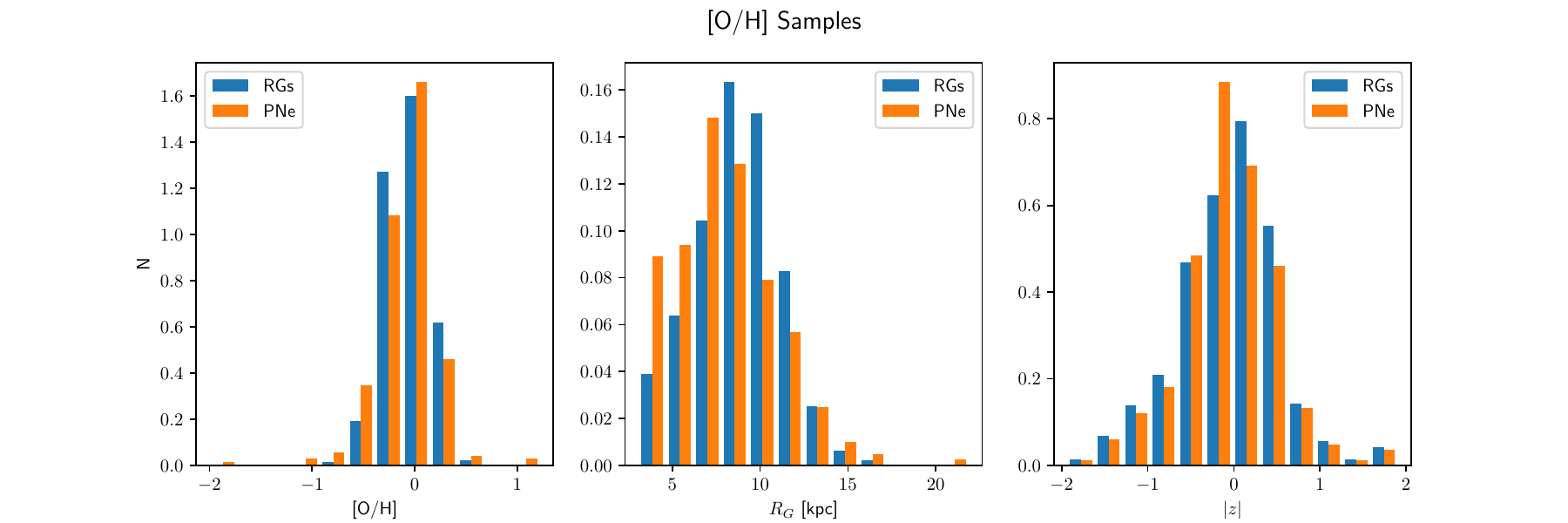}
      \caption{The distributions of [O/H] (left panel), $R_{\rm G}$ (central panel), and $z$ (right panel) in the [O/H] PN and RG samples (see selection in the text).}
      \label{3distr}
    \end{figure}

To perform our analytic comparison between RGs and PNe, for each reference PN sample (e.g., the [O/H] PN sample, the [Fe/H] PN sample, etc) we build RG samples starting from our initial RG sample, but whose spatial and metallicity are distributed similarly to the reference PN sample. This step is necessary to compare population pairs directly. 

We test the parameter distributions of the RG samples by running the Kolgomorov-Smirnov (KS) test between the distribution of the spatial parameters and metallicity of each RG sample with the corresponding PNe sample. If the result rejects the null hypothesis (i.e., p$<$0.05), we exclude the newly created RG sample. Any RG sample that is not excluded could be used for the subsequent analysis, as discussed also in Section~\ref{sec:discussion}. 

For example, we can build N [O/H] RG samples,  where N ([O/H] PN sample) $<$ N $<$ N (initial RG sample), and reject N$_{\rm rej}$ of them based on the KS test results. We then randomly choose the RG [O/H] sample of Table~\ref{KS} among the N - N$_{\rm rej}$  possible [O/H] RG samples, and the results would not change had we selected another [O/H] RG sample. Then, we repeated the procedure for the other [X/H] samples.

\begin{table}
\caption{Comparison of PN and RG distributions}          
\label{KS}     
\scriptsize
\begin{tabular}{llllll}
\hline
{Samples}&  {N$_{\rm PNe}$}&  {N$_{\rm RGs}$}& parameter/element&  {Med (PNe)}& {Med (RG)} \\
\hline

PNe, RGs, [O/H] samples&    258& 28,514& & & \\
&&&$R_{\rm G}$& 7.628&7.879\\
&&&    $z$&      -0.074& -0.022\\
&&&     	[O/H] &   -0.060& -0.062\\
&&&&&\\
\hline
PNe, RGs, [Fe/H] samples&     	33& 25,065&& & \\
&&&$R_{\rm G}$&   7.261& 7.655\\
&&&$z$&   0.166& 0.221\\
&&& [O/H]&   -0.020&-0.029\\
&&&& & \\
\hline
PNe, RGs, [S/H] samples&     	31& 21,830&& & \\
&&&$R_{\rm G}$&   7.590& 8.123\\
&&&$z$&   -0.007& 0.016\\
&&&	[O/H]&  -0.058& -0.049\\
&&&& & \\
\hline
PNe, RGs, [N/H] samples&    246& 28,873&& & \\
&&&$R_{\rm G}$&  7.608 &7.841\\
&&& 	$z$& -0.070& -0.028\\
&&&[O/H] &  -0.060 &-0.060 \\
&&&& & \\
\hline
PNe, RGs, HBB samples&    22& 28,873&& & \\
&&&$R_{\rm G}$&  7.11 &7.841\\
&&& 	$z$& -0.071& -0.028\\
&&&[O/H] &  -0.045& -0.06\\
&&&& & \\
\hline
PNe, RGs, [C/H] samples&  55& 26,074&& & \\
&&&$R_{\rm G}$&  8.092& 8.272\\
&&& 	$z$&  -0.054&-0.005\\
&&&[O/H] &  -0.050&-0.046\\
&&&& & \\
\hline

\hline

\end{tabular}
\end{table}

 \section{Metal Indicators and Biases}\label{sec:metal}

The abundance ratios of $\alpha$-elements over iron have been broadly used in Galactic Archaeology as these can provide important information about the chemical evolution of the Galaxy, with the $\alpha$-elements, such as oxygen, being produced in short timescales (a few Myr) by SNe II and iron being mostly produced on considerably longer timescales (hundreds of Myr) in SN Ia. This pioneering idea of \citet{Tinsley1979} has been widely accepted since then, and the timescales for the explosions of SN Ia are now empirically estimated through observations of their Delayed Time Distribution  in extragalactic systems, e.g. \citet{Maoz2017}; see also Figure C1 in \citet{Kubryk2015}.
The [$\alpha$/Fe] ratio stays constant with increasing [Fe/H] until the SN Ia evolution onset in the Galaxy, which makes the [$\alpha$/Fe] ratio decrease \citep[see e.g. Fig. 13 in][]{Pr+2018}.  

Both $\alpha$-elements and [Fe/H] can be used as metallicity indicators in stars. Iron is the reference metallicity element observed in stars. APOGEE DR~17 provides iron abundances derived from \ion{Fe}{1} lines, and metallicities derived from overall fits of the APOGEE spectra, with these two estimates producing very similar metallicity values. Iron is observed in emission lines of some PNe, but most iron in PNe is in the solid and not the gaseous state, thus can not be used as metallicity indicator. We explore the depletion of iron (and sulfur) in Section~\ref{sec:depletion}. 

Oxygen is the most abundant element after H and He and can be used as a metallicity indicator both in RG stars and in PNe if we assume that the oxygen abundance variation in the advanced stages of stellar evolution is minimal in the observed stars and nebulae. Oxygen has the observational advantage that is measurable both in the stellar sample and in PNe. Stellar evolution models indicate that oxygen abundance is affected by both the DU episodes and the HBB.
\citet[][]{Ventura15} indicate that the final enrichment of oxygen in stars with initial mass $0.85< M < 3.5 $\msun is due to DU processes, and it depends on the initial stellar mass. Furthermore, while the DU event per se does not depend on metallicity, the final enrichment measured as the [O/Fe] ratio is higher for low metallicity stars since the initial [Fe/H] is lower at low metallicity. While oxygen enrichment theoretically can be as high as 0.6 dex for the highest masses in this low-mass range ($M\sim$ 3.5\msun) and the lowest initial metallicity, an observed sample of PNe whose progenitor masses are lower than $\approx3.5$\msun and follow the IMF favoring low mass stars, does not typically suffer from enrichment greater than 0.1 dex, which is lower than the typical observational uncertainties. 

For PNe whose progenitor are in the higher mass range there is no oxygen enrichment in the AGB, due to the onset of the HBB, although the HBB is related to oxygen depletion \citep{Ventura15}, another issue to consider when using oxygen as a metal indicator. While in principle all stars with mass $M\gtrapprox$ 1\msun may evolve through the AGB phase, the {\it observed} PNe typically have progenitors mass $M\lesssim 3 - 5$\msun, depending on metallicity.  The principal cause of the low observation probability of PNe with high-mass progenitors derives from the shape of the IMF. If we use Salpeter's IMF for the Galactic PN progenitors, only 4.5$\%$ of the progenitors would have masses above 5\msun, and less than 14$\%$ would have masses above 3\msun. This low percentage of high-end mass LIMS is further reduced in observed PN statistics by the PN lifetime, which crucially depends on the CS mass, which, in turn, is related to the initial mass. High-mass CSs evolve so quickly toward the low luminosities in the post-AGB phase that their PNe are rarely observed. A complete population synthesis of post-AGB stars and PNe with the most recent models and observational input would be required to quantify the expectancy of observing PNe with high progenitor masses. Past population synthesis studies, limited to a narrow metallicity range \citep[][e.~g.]{2001A&A...378..958M, 2000ApJ...542..308S} are useful indicators of these expectations.

\citet{2021AJ....161..183M} compared chlorine and oxygen abundances derived for a sample of $\sim$50 RG stars with literature abundance results for PNe from \cite{DIRP15} and Henry et al. (2004), finding good agreement, which would indicate that both chlorine and oxygen in PNe are good metallicity probes for {\it observed} PN samples.  \citet{2023MNRAS.519.2169K} published a comprehensive table of oxygen enrichment in AGB stars by mass and metallicity, finding that one expects no enrichment in $M\lessapprox 1$\msun stars at any initial metallicity, and that the oxygen enrichment expected in {\it observed} PNe with low progenitor mass is comparable to the observational uncertainties since most of the observed PNe would be the progeny of stars with masses very close to 1\msun, due to the shape of the IMF.  Thus, the expectancy from evolutionary models indicates that the yield of O$^{16}$ in observed LIMS remnants at solar metallicity is very low or null; nonetheless, there are exceptions in the literature, such as selected carbon-rich dust PNe \citep{DIRP15}, which would make these particular PNe not ideal probes for Galactic chemical evolution. 

We use oxygen as the metallicity indicator through the paper, where the [O/H] measured in PNe is used to build the RG samples, together with the spatial parameters.

\section{Iron and Sulfur Depletion}\label{sec:depletion}

The formation of solid particles (dust) takes place in the cool outer layers of highly evolved RGs, which are then ejected during the PN stage
and become the material expanding into the ISM, with gas-phase depletion now imprinted onto the PN gaseous abundance distributions. \citet{KL2014} point out that the LIMS may contribute as much as $\sim$90\% of the dust in the ISM.
The effective temperatures of RG stars considered here are all hotter than $T_{\rm eff}>3500$ K, where dust formation in the spectral-line forming photospheric layers (with $T\sim$2900-3200 K) is negligible, as dust formation occurs only in the outermost layers of these giants, where temperatures fall to $T\sim$1500-1700 K \citep[e.~g.,][]{T2022,L2003}, thus the photospheric abundances represent the true chemical composition of the star.

\begin{figure}
   \centering
   \includegraphics[width=12cm]{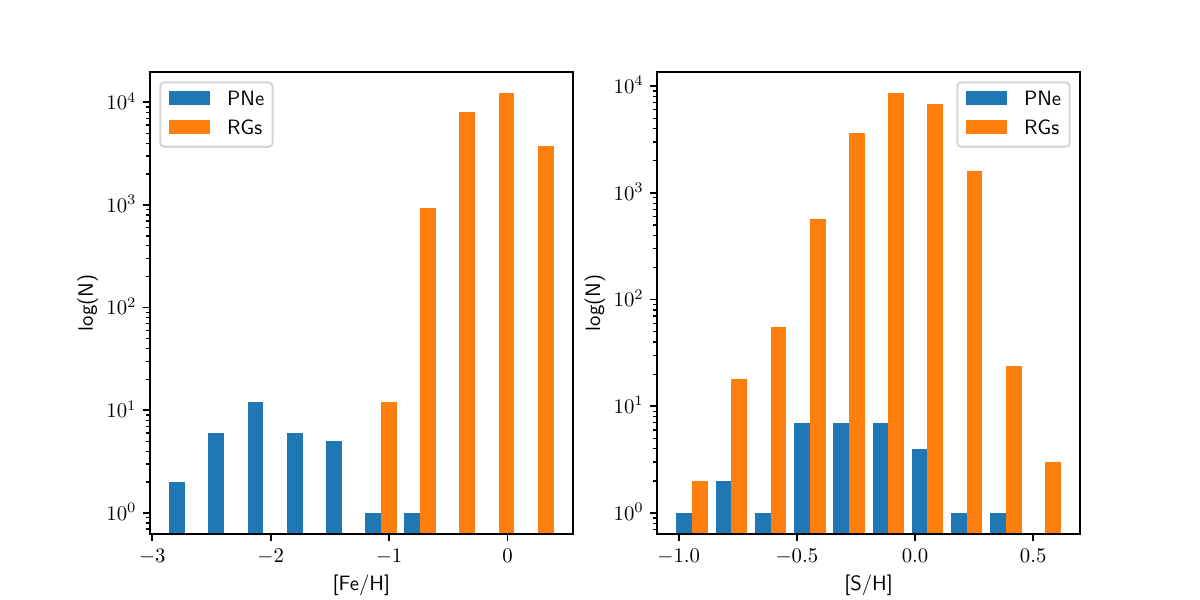}
      \caption{Left: [Fe/H] distributions in [Fe/H] PN and RG samples. Right:  [S/H] distributions in [S/H] PN and RG samples}
         \label{depletions}
   \end{figure}
   
Dust in PN gas, on the other hand, is essential in the mechanism of envelope ejection, which is mainly due to pressure on the dust grains
\citep{1974IAUS...66...62P}, and its composition depends on the initial mass and metallicity of the progenitor \citep[e.~g.,][]{Ventura15}.  
The amount of depletion suffered by a particular element is a function of its condensation temperature, $T_{\rm c}$, where $T_{\rm c}$ is defined as the temperature at which 1/2 of the atoms of the element in-question will be condensed into solids for a given chemical composition and gas pressure.  \citet{L2003} provides condensation temperatures for all elements using a solar composition at $P=10^{-4}$ bar and, as a point-of-reference, the values of $T_{\rm c}$ for the elements discussed here are $T_{\rm c}$= 77 K for C, 131 K for N, 181 K for O, 693 K for S, and 1351 K for Fe.  Although this is for solar composition, the order of the values of $T_{\rm c}$ will change only for extreme chemical compositions. Depletion is expected to increase for elements with high values of $T_{\rm c}$ (refractory elements) and to be small or negligible for low values of $T_{\rm c}$ (volatile elements).  Since PN gas consists of material that originated in the outer envelopes of cool, luminous RGs, where dust formation will have occurred, it is necessary to consider depletion effects in the abundance comparisons.

Based on their relative condensation temperatures, it would be expected that Fe would suffer from more significant depletion than the other elements considered here and, indeed, iron depletion has been inferred previously by comparing PN abundances to solar photospheric abundances \citep{DIR14}, with the depletion explained as due to condensation of this element onto dust grains. In this study, the large depletion of iron is obvious when comparing the distributions of the values of [Fe/H] in the RG and PNe as shown in the left panel of Figure \ref{depletions}. In this panel we see qualitatively that the two distributions differ especially toward the higher [Fe/H] values.

The iron depletion, hereafter D[Fe/H], is generally defined as the difference between the solar iron abundance and that of the gas-phase iron abundance of the nebular probe \citep{DIR14}, allowing for a quantitative measurement of the effect. Nonetheless, iron depletion in PNe has not yet been measured relative to the underlying stellar population of comparable oxygen abundance to the PN progenitor. This is what is done here, with the aim of characterizing iron depletion relative to the initial stellar metallicity.

\begin{figure}
  \centering
    \includegraphics[width=8cm]{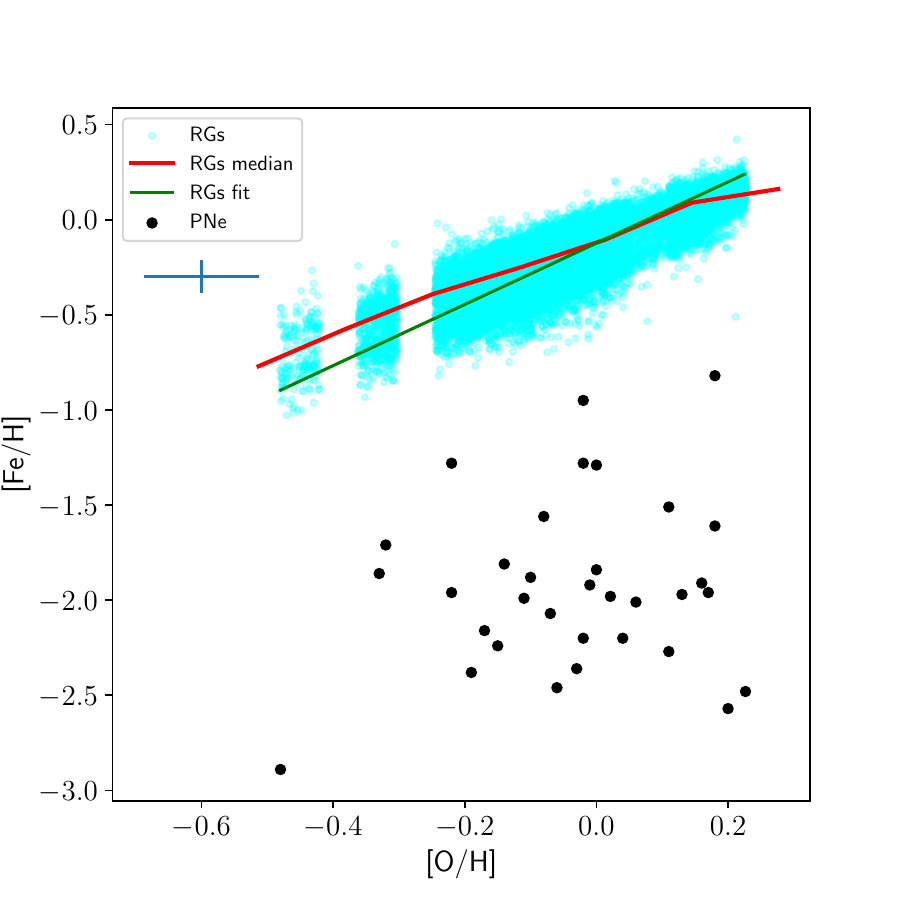}
  \caption{[Fe/H] vs. [O/H] of [Fe/H] RG (cyan circles) and PN (black dots) samples. The blue cross is the assumed Pn abundance errorbar (see text). We show the running median of the RGs, and the [Fe/H] vs. [O/H] linear fit, see text. The iron deficit in PNe is evident across the population. }
       \label{iron_dep1}
   \end{figure}

In Figure~\ref{iron_dep1} we plot the [Fe/H] PNe and RG samples in the  [Fe/H] versus [O/H] plane. The [Fe/H] PN sample consists of all PNe with at least one [Fe/H] measurement. The [Fe/H] RG sample has spatial and [O/H] distribution compatible with those of the [Fe/H] PN sample. Sample sizes and KS statistics are in Table~\ref{KS}. In the Figure we also indicate both the running median of the RG distribution and the linear fit to the RG abundances, [Fe/H]$_{\rm RG}$=(1.610$\pm$0.00543) [O/H] - (0.123$\pm$ 0.000746). The PN sample with measured iron abundances spans a relatively small range in oxygen abundances around solar. 

We measure the iron depletion of the PNe as:

$${\rm D[Fe/H] = | [Fe/H]_{\rm RG} - [Fe/H]_{PN}}|  ,\eqno(1)$$

where [Fe/H]$_{\rm RG}$ has been estimated from the linear fit of the [Fe/H] versus [O/H] for each PN, and not simply by scaling to the solar value of [Fe/H]. 
We find that the average iron depletion is $<$D[Fe/H]$>$=1.741$\pm$0.486.

 \begin{figure}
   \centering
   \includegraphics[width=8cm]{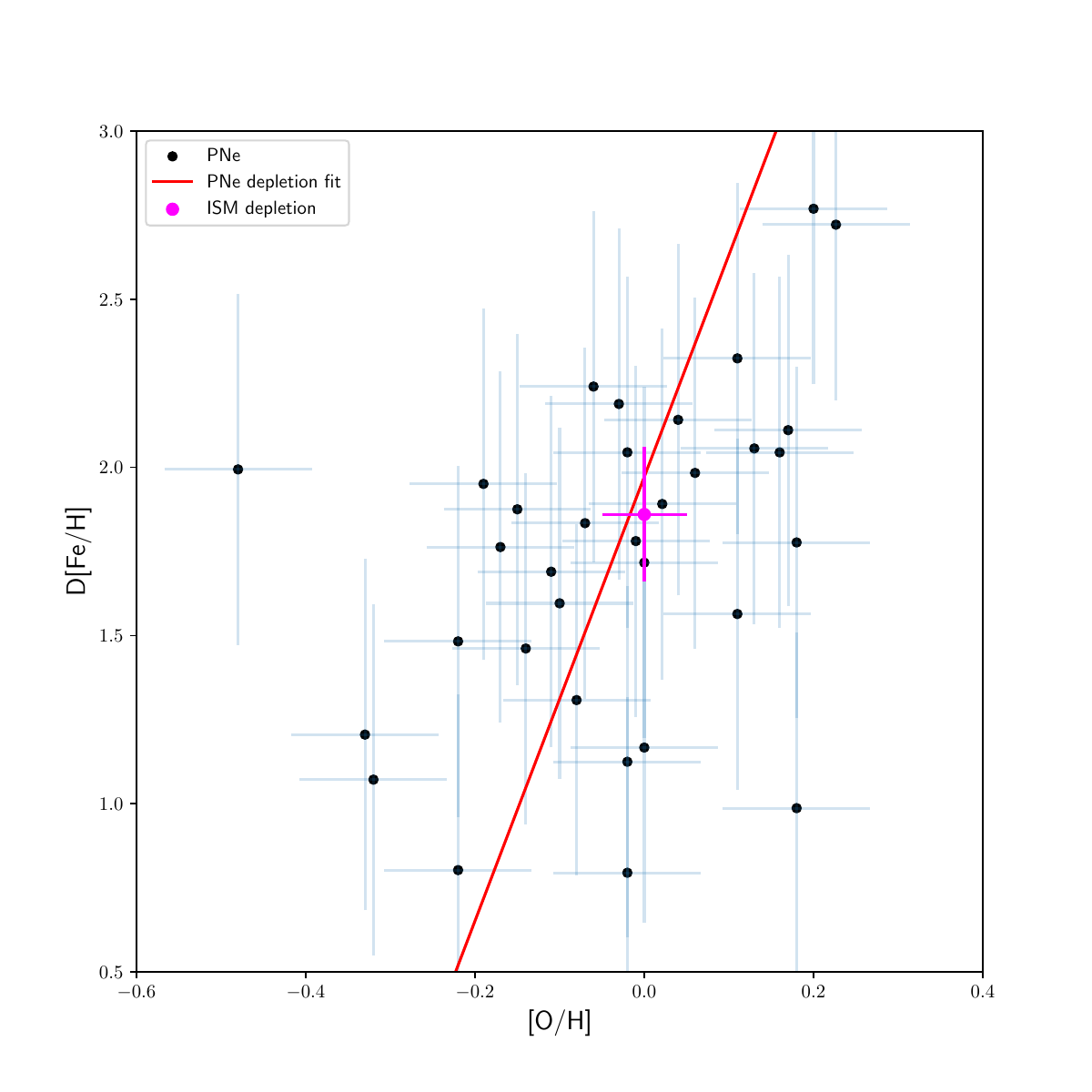}
      \caption{The iron depletion of the PNe in our study, D[Fe/H], vs. [O/H].  The red line indicates the linear fit. The magenta dot and errorbar correspond to the ISM value for D[Fe/H], as described in the text.}
         \label{iron_dep2}
   \end{figure}
  
In Figure~\ref{iron_dep2} we show D[Fe/H] as a function of [O/H].  We find a positive, mild correlation between D[Fe/H] and [O/H], with linear fit of the logarithmic abundances D[Fe/H]$=(6.6003\pm2.443)\times{\rm [O/H]} +(1.972\pm0.199)$. The Spearman correlation coefficient using typical uncertainties is 0.12$\pm$0.028.
 
  At a given value of [O/H] in Figure~\ref{iron_dep1}, the range in [Fe/H] spans approximately $\pm$0.5 dex for the RG stars.  Across a wide range of abundances for Fe of $\approx$2.5 dex (a factor of 300), the scatter in [Fe/H] as a function of [O/H] does not vary significantly.
The linear fit to the RG points in the figure reveals the underlying signature of the Galactic chemical evolution of both O and Fe; at [O/H]=+0.5 the value 
of [Fe/H]=+0.682, while at [O/H]=-0.5 the value of [Fe/H]=-0.928.  These mean values are defined by the linear fit result in [O/Fe]=-0.182 at [O/H]=+0.5, which increases to [O/Fe]=+0.428 at [O/H]=-0.5. The increasing value of [O/Fe] with decreasing metallicity typifies the increase in the oxygen abundance relative to iron at lower metallicities in both the Galactic thin and thick disks. 

The behavior of [Fe/H] with [O/H] in the PNe is different and not orderly, with a significant spread in [Fe/H] over a very small range in [O/H], with the resultant values of [O/Fe] as large as +2.5.  As the PN progenitor stars are not expected to be significant sources of O, nor sinks of Fe (via nuclear processes), the extreme values of [O/Fe] in the PN of Figure~\ref{iron_dep1} are not due to nuclear reactions or Galactic chemical evolution, but are likely due to chemical processes, i.e., condensation of Fe into a non-gas-phase (solid) component. 
  
 The large iron depletion factor found here can be compared to [Fe/H] abundances measured along lines-of-sight through diffuse ISM gas, as collected in \citet{ZLZ2021}, who present [Fe/H] abundances from 50 lines-of-sight from the following studies: \citet{JSS1986}, \citet{VSS1988}, \citet{SRF2002}, \citet{SFW2003}, \citet{MLS2007}, \citet{JS2007}.  These lines-of-sight sample the diffuse ISM towards relatively bright stars that are at distances of $\approx$200 pc out to a few kpc from the Sun, with median and MAD values for this sample of [Fe/H]=-1.86$\pm$0.20.  This interstellar Fe abundance represents the current ISM and falls quite close to the average value of D[Fe/H] derived for the PN sample discussed here.  The agreement in the magnitude of the iron depletion in both the PN gas and ISM gas illustrates the close connection between the chemistry of PNe and the ISM.  It is also worth noting that the analysis techniques used for PNe are quite different from those used to analyze interstellar absorption lines. 

The average D[Fe/H] is also compatible with other values found in the literature, which are  all measured with respect to solar \citep{DIR14}. \citet{KH22} presented a thorough summary of the various iron depletion studies available in the literature, including results by \citet{Shields1978} and \citet{Garstang1978}. Solar photospheric abundances could be considered ideal benchmarks for inferring average depletions at roughly solar metallicities. In this study we have the opportunity to go a step further and estimate depletions relative to a stellar population that covers a range in [O/H] and not to refer to a single reference star.

We investigated D[Fe/H] as a function of planetary nebula parameters, such as morphology, angular and linear radius, dust type, ionized mass, central star temperature, distance from the Galactic center and the Galactic plane, nebular Balmer flux and extinction, and nebular surface brightness, using parameters collections in \citet{SH18} and \citet{BS23}, but found no clear correlations between D[Fe/H] to any of these parameters. The only clear correlation remains that D[Fe/H] is larger for higher values of [O/H].

Sulfur, with a condensation temperature of 693 K \citep{L2003}, falls between the low-$T_{\rm c}$ volatile elements C, N, and O (with $T_{\rm c}$=77 K, 131 K, and 181 K, respectively) and the high-$T_{\rm c}$ refractory element Fe ($T_{\rm c}$=1351 K), and possible S abundance depletion factors in PN gas remain uncertain. Evidence for a deficit in the sulfur abundances of PNe has been previously investigated \citep{KH22}. Sulfur depletion is defined here as: 
$$ {\rm D[S/H] = | [S/H]_{\rm RG} - [S/H]_{PN}} |. \eqno(2)$$

Sulfur in PNe is measured from the strong [\ion{S}{2}] emission lines at $\lambda$6716~\AA~and $\lambda$6731~\AA, as well as the [\ion{S}{3}] $\lambda$6312~\AA~line, and/or the IR lines, such as the [\ion{S}{3}] line at $\lambda$9532~\AA. In cases where S$^{2+}$ emission is not measured, ICFs are computed from fits to other ionic ratios, such as O$^{+}$/O, in order to estimate the total sulfur abundance \citep{KB94}. \citet{HSK12} examined the possible D[Fe/H] in PNe and, based on results for H~II regions, tentatively concluded that the deficit may result from inaccurate ICFs, noting that D[Fe/H] is lower -- albeit not eliminated -- when the abundance uses measured [\ion{S}{3}] lines. The dependence of D[Fe/H] from the presence of the sulfur's upper ionized state, as pointed out by \citet{HSK12}, led us to discover a possible cause for artificial deficit, not due to any astrophysical reason, which is presented in the Appendix, and used in the following sulfur analysis.

     \begin{figure}
   \centering
   \includegraphics[width=8cm]{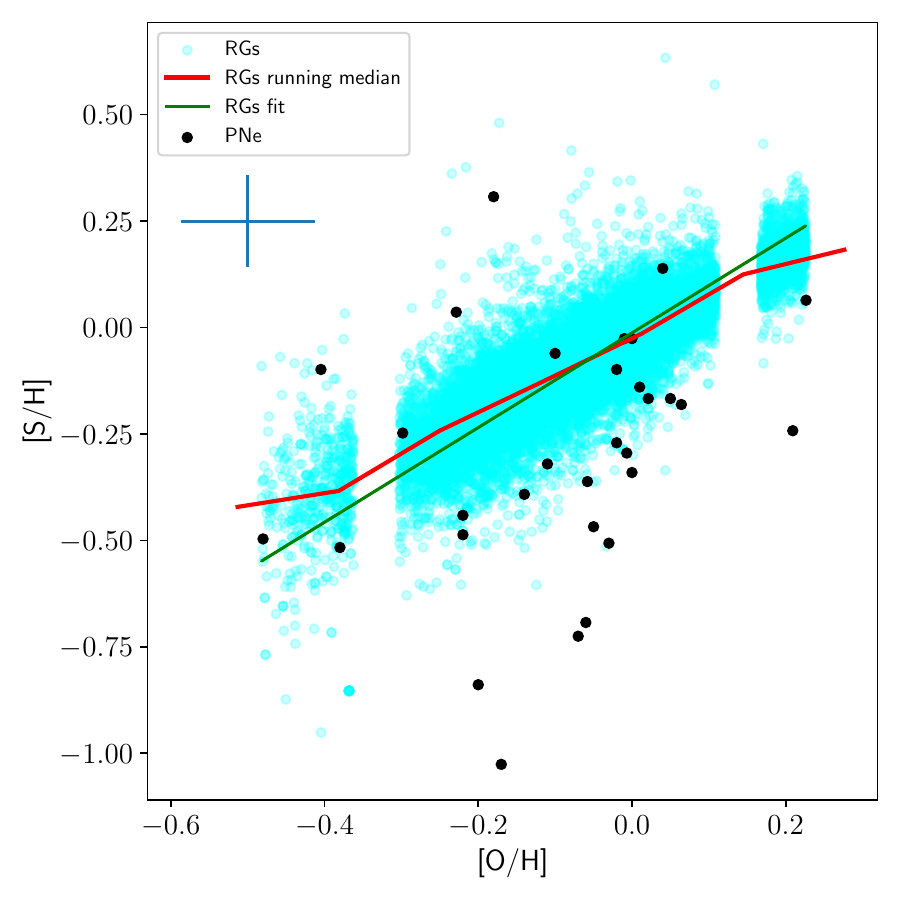}
      \caption{As in Fig.~\ref{iron_dep1}, but for [S/H] vs. [O/H], using the [S/H] PN and RG samples. 
       }
         \label{sul_dep1}
   \end{figure}

The presence, or lack thereof, of sulfur depletion in PN gas remains an open question. In Figure~\ref{depletions}, right panel, we compare [S/H] in PNe and RGs (sample sizes and KS statistics regarding their spatial and [O/H] distributions of the [S/H] sample are in Table~\ref{KS}), and we can see mild sulfur depletion in the PN population. We examine the sulfur vs. oxygen abundance, just like we did above for iron.

In Figure~\ref{sul_dep1} we show the loci of the [S/H] PN and RG samples in the [S/H] vs. [O/H] plane. The [S/H] PN sample includes all PNe with [S/H] measurements, while the [S/H] RG sample has been selected based on the [S/H] PN sample for the spatial and metal distribution, as in the [Fe/H] samples. We exclude sulfur abundance results that use Eq.~38A in \citet{KB94} (see the discussion in the Appendix). We found a sulfur deficit in PNe, with an average $<$D[S/H]$>=0.179\pm0.291$ dex measured from the linear fit of the RGs to the PN vertically, for the [O/H] value of the PN, as in the [Fe/H] case. The average depletion of sulfur is lower than the average D[Fe/H], but still significant. 

There is no clear correlation between D[S/H] and the [O/H] abundance, and we did not find any other significant trends between D[S/H] and other nebular characteristics.  

D[S/H] for the PNe in our sample is in reasonable agreement with the trend found in the ISM, as summarized in \citet{RJT2022} in their Figure 5. The D[S/H] value for the ISM at solar metallicity, D[S/H]$_{\rm ISM}$=0.45$\pm$0.28, is compatible with the average PNe depletion within the uncertainties. It is worth noting that the comparison of sulfur abundances in the Orion B stars and the Orion nebula also lead to the conclusion that sulfur depletion into grains is from modest to non-existent \citep{DCR2009}. 

Given the agreement between what we find here and the results from the ISM, we conclude that PNe exhibit mild sulfur depletion. 

\section{Carbon and Nitrogen enrichment in PNe} \label{sec:evolution}

LIMS  (i.e., stars that do not become supernovae of Type II) are major sources of carbon and nitrogen in the Universe, with mass integrated yields rivaling those of the high mass stars. 

\citet{KL2014} provide a detailed review of the various stages and nucleosynthesis yields from the LIMS as they evolve along the RGB and the AGB. According to the Galactic chemical evolution model of \citet{P+2023}, which includes yields from rotating massive stars, LIMS produce about 1/3 of C, while massive stars produce the remaining fraction. They also find that a large amount of N, $\approx$45\%,  originates in LIMS. However, \citet{P+2023} notice that the LIMS contribution may be revised downwards for C and upwards for N, if HBB (not included in their model) is taken into account.
\citet{2023arXiv230207255K} give yields of C and N contributions through Galaxy evolution from both the massive stars and LIMS, including the HBB processing, and found that 49$\%$ of carbon and 74$\%$ of nitrogen originates in LIMS nucleosynthesis. While the final percentages are still under debate across models, there is certainty that the LIMS are major players in the production of these elements in the Universe.

\begin{figure}
   \centering
   \includegraphics[width=12cm]{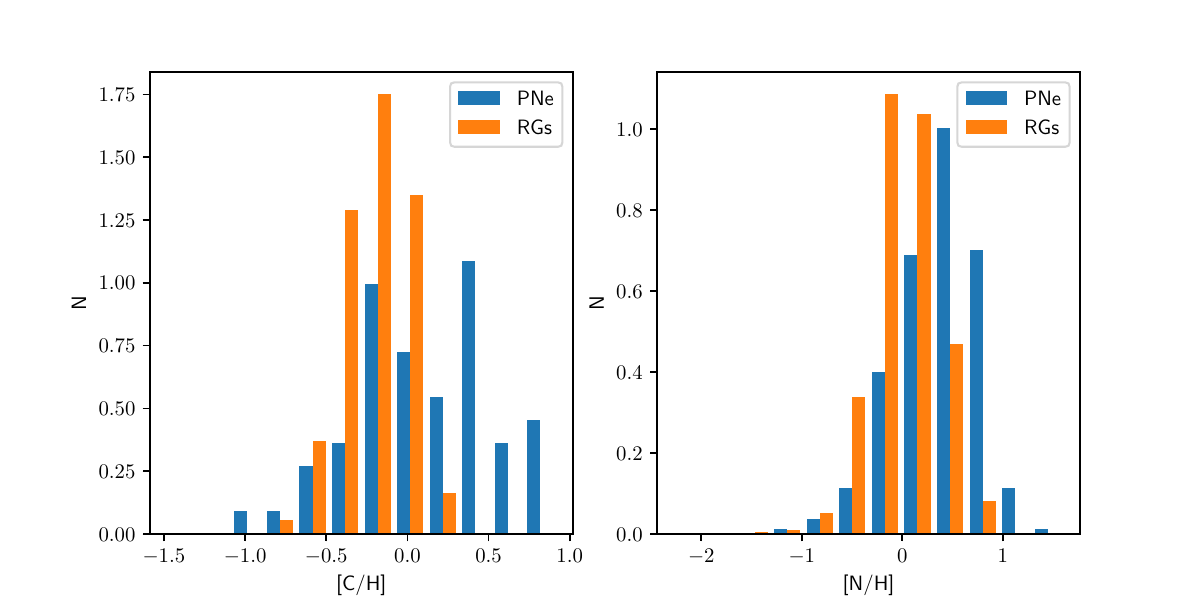}
      \caption{The probability density of [C/H] (left panel), and [N/H] (right panel) in the [C/H] (left panel) and [N/H] (right panel) PNe and RG samples.}
         \label{NHCH}
   \end{figure}

\begin{figure}
   \centering
   \includegraphics[width=12cm]{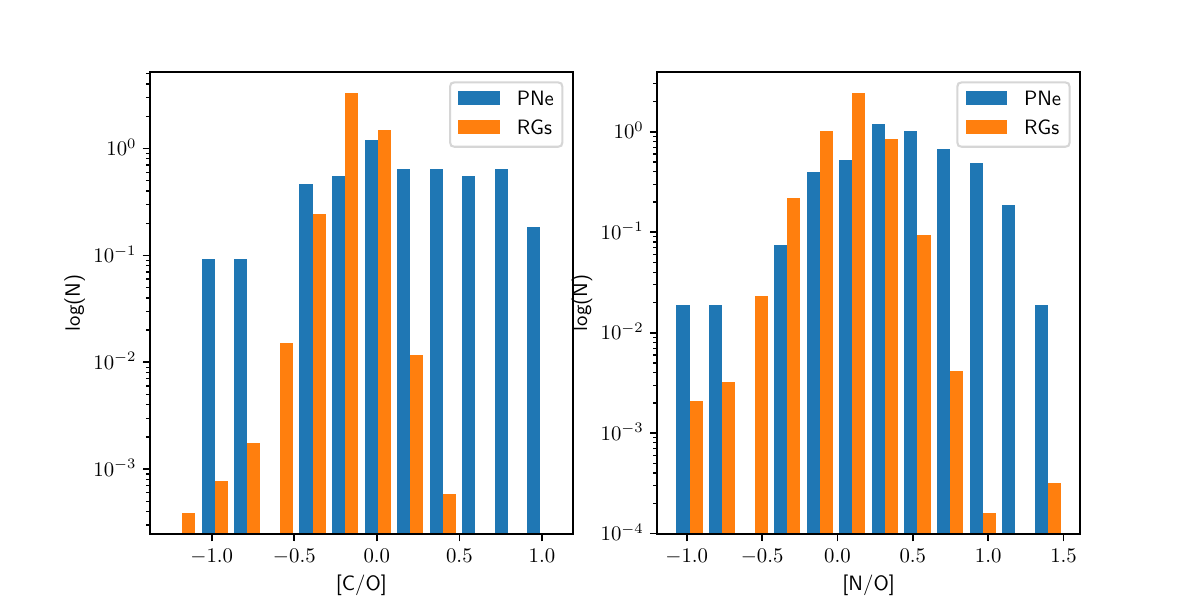}
      \caption{Same as in Fig.~\ref{NHCH}, but for [C/O] (left panel) an d[N/O] (right panel).}
         \label{NOCO}
   \end{figure}
      
In Figure~\ref{NHCH} we show the [C/H] and [N/H] abundance distributions of the corresponding PN and RG samples, selected as for the other elements.  The compared  [C/H] and [N/H] nebular and stellar samples are described in Table~\ref{KS}.  The PNe contain significant fractions of C-rich and N-rich examples. Figure~\ref{NOCO} removes the stellar metallicity and shows the abundance ratios of [C/O] and [N/O] for the RG stars and PNe. The [C/O] and [N/O] ratios reveal the same trend as the [C/H] and [N/H] abundances; the PNe contain a significant fraction of very C- and N-rich, relative to O, objects.  The excess carbon and nitrogen abundances originate from triple-$\alpha$ burning and 3$^{\rm rd}$ DU of primary $^{12}$C during the TP-AGB phase of evolution, along with the conversion of some of this $^{12}$C into $^{14}$N in the more massive IMS during HBB.  
   
PNe represent the final stage in the evolution of LIMS during their transition from high-luminosity stars to white dwarfs, and they capture a snapshot of this final evolution of the carbon and nitrogen abundances.  This final stage of stellar evolution is not well-captured in the stellar sample, which is dominated by less-evolved oxygen-rich RGs.  Carbon stars are part of the APOGEE sample, however, as can be seen in the left panel of Figure~\ref{NOCO}: values of [C/O]$\gtrapprox +$0.3 have ratios of C/O$\gtrapprox $1, and thus represent the carbon stars in the APOGEE sample.  The number of carbon stars in the left panel of Figure~\ref{NOCO} relative to the oxygen-rich giants (with C/O$<$1) is $\approx3.5\times10^{-4}$.  This can be compared to estimates of the space density of carbon stars, relative to GKM giants, in the Milky Way, $\approx10^{-3}$  \citep{1987ApJS...65..385C, 2004A&A...418...67N}
with uncertainty in this ratio of at least a factor of a few.  It is also expected that a NIR H-band survey of stars will miss a fraction of the very evolved carbon stars, as at the very final stage of their evolution, carbon stars become heavily enshrouded in dusty, C-rich envelopes due to extreme mass-loss rates as they evolve towards the PN phase of evolution 
\citep[e.~g.][]{2021csss.confE.165S, 2019ApJ...887...82K}.

\begin{figure}
   \centering
   \includegraphics[width=8cm]{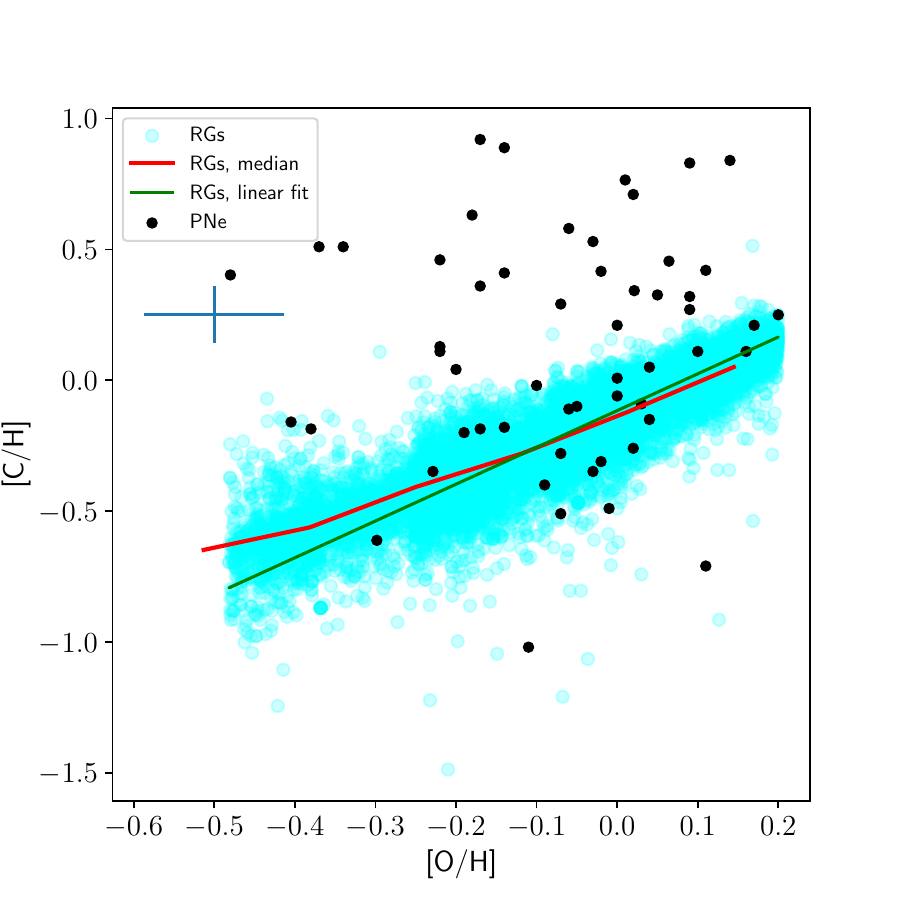}
      \caption{As in Fig.~\ref{iron_dep1}, but for [C/H] vs. [O/H], using the [C/H] PN and RG samples. 
. }
         \label{CH_enrich}
   \end{figure}

We study carbon and nitrogen enrichment in PNe with the same method that we have used for the iron and sulfur depletion. 
Carbon enrichment is shown in Figure~\ref{CH_enrich}, where we plotted the [C/H] PN and RG samples. We measure the enrichment as:
$${\rm E[C/H] = | [C/H]_{\rm RG} - [N/H]_{PN}} |, \eqno(3)$$
and in Figure~\ref{CH_enrich} we see the evidence that the majority of the observed PNe have enhanced carbon against the underlying progenitor population. 

The RG sample, fitted with the usual method, is used in Eq. (3) to find [C/H]$_{\rm RG}$=(1.406$\pm$0.00359)$\times$[O/H]-(0.116$\pm$0.000558), and the average E[C/H] in PNe, from Eq. (4), is $<$E[C/H]$>$=0.337$\pm$0.463.

\begin{figure}
   \centering
   \includegraphics[width=8cm]{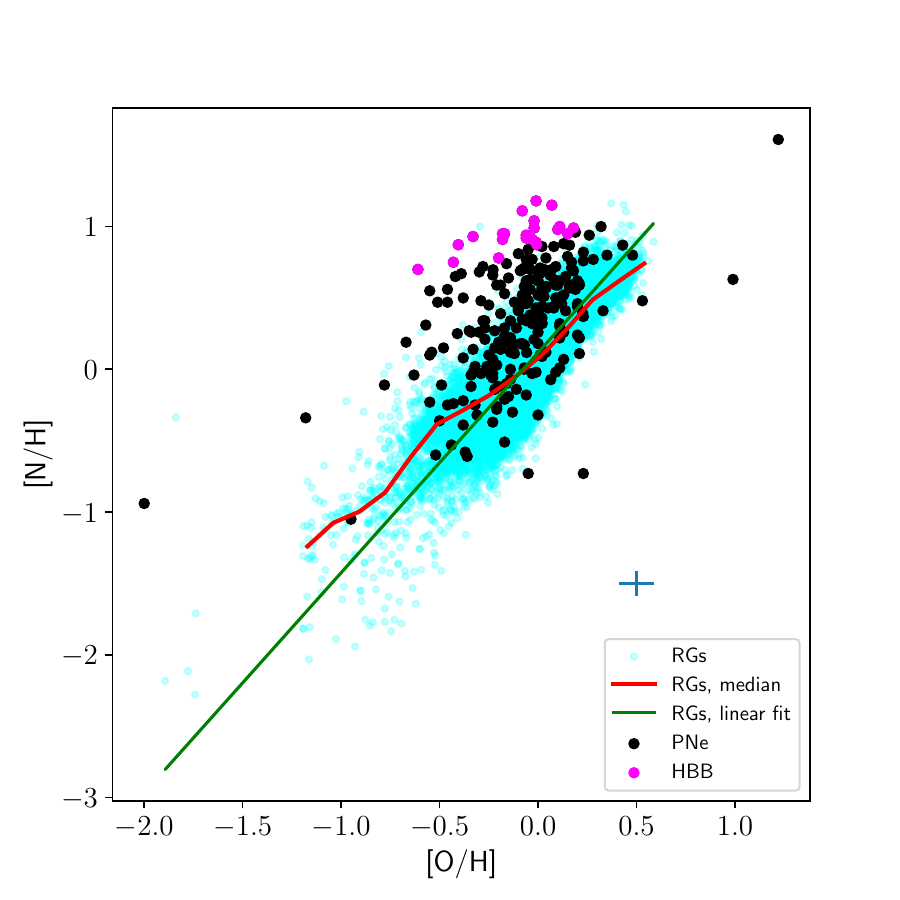}
      \caption{As in Fig.~\ref{iron_dep1}, but for [N/H] vs. [O/H], using the [N/H] PN and RG samples. 
}
         \label{NH_enrich}
   \end{figure}

Nitrogen enrichment is shown in Figure~\ref{NH_enrich}, where [N/H] has been plotted against [O/H] for the [N/H] PNe and RG samples (see Table~\ref{KS}). The fit to the RG relation is [N/H]$_{\rm RG}$=(1.542$\pm$0.00508)[O/H]+(0.117$\pm$0.000184). We define PN nitrogen enrichment against the RGs at the [O/H] PN value, similar to iron and sulfur depletion:

$${\rm E[N/H] = | [N/H]_{\rm RG} - [N/H]_{PN}} |, \eqno(4)$$

and found that the average [N/H] enrichment in PNe is $<$E[N/H]$>$=0.393$\pm$0.421.

 It is worth examining the above result in the light of the PN populations whose progenitors go through the HBB processing.  Figure 1 in \citet{2017MNRAS.471.4648V}, right panel, shows the snapshot of the final yields of stellar evolution of LIMS for a broad selection of initial metallicity, describing the footprint of the PN abundances in the log(N/H)+12 versus log(O/H)+12 plane. By using the updated models of \citet{2017MNRAS.471.4648V} (see also \citet{2023MNRAS.519.2169K}, we determine that PNe with
$ {\rm [N/H]} > 0.5\times{\rm [O/H]} + 0.870$ evolve though the HBB. Following the initial masses and the ages of the evolutionary models, we can say indicatively that the progenitor ages of HBB PNe progenitors are younger than $\approx$1 Gyr. 

In Figure~\ref{NH_enrich} we indicate which PN progenitors experienced the HBB. About 10$\%$ of the [N/H] PN sample is HBB-processed. We compare these PNe with the [N/H] RG Sample, and give the results in Table~\ref{KS}.  The [N/H] enrichment measured for the HBB PNe with respect to the RGs is higher than that for the whole PN sample, 
E[N/H] = | [N/H]$_{\rm RG}$ - [N/H]$_{\rm HBB PNe}$ |=0.980$\pm$0.243. 
This is near the upper limit of the PN [N/H] enrichment.

There are only 2 HBB RGs in the sample of Figure~\ref{NH_enrich}, in the RG sample of 28,873 stars; excluding these two stars from the RG fit would not change the result notably.

\section{Discussion and Concluding Remarks}\label{sec:discussion}

This study is based on the assumption that Galactic PNe are the progeny of Galactic RG stars. We compared selected samples of PNe and RGs whose spatial and metallicity ([i.e., [O/H]) distribution are are consistent with not being different, and then examine the elements that we know could be depleted by condensation, or enriched by AGB evolution. 

We found that PNe are iron and sulfur depelted. We confirmed that most of the iron in PNe is entrapped in grains, with $<$D[Fe/H]$>$=1.741$\pm$0.486 dex. We also found that the iron depletion mildly correlates with [O/H]. If [O/H] can be used as a metallicity indicator in PNe -- and we argue that it could, given the narrow distribution in mass of PN progenitors -- we conclude that iron depletion weakly depends on metallicity. 
Sulfur depletion is milder,  $<$D[S/H]$>=0.179\pm0.291$ dex. The large error bar makes the conclusion tentative.

We studied the elements that vary with the AGB evolution, such as carbon and nitrogen, and we quantified average nitrogen enrichment, $<$E[N/H]$>$=0.393$\pm$0.421 dex,  although the large uncertainty making the conclusion tentative. The enrichment  doubles if we limit the analysis to those PNe whose progenitors have gone through HBB processing. We also found average carbon enrichment,  $<$E[C/H]$>$=0.332$\pm$0.460 dex, also with a large uncertainty.

 Through the paper we selected large RG samples for our analysis (see Table~\ref{KS}). We have tested whether the sample size affect the analysis by repeating the whole depletion and enrichment measurements with RG samples of different sizes, and we obtained elemental depletions and enrichments within 0.01 dex of those given in this paper. As an example, by selecting a small (N=246) RG sample, whose Galactic radius and altitude and [O/H] distributions were selected based on the [Fe/H] PN sample with p$_{\rm KS}>0.05$ for the spatial and metallicity sample comparisons, we found $<$D[Fe/H]$>$=1.733$\pm$0.488 dex.

PNe have been used in the extragalactic context to set constraints to the chemical evolution of galaxies for quite a long time \citep[e.~g.][]{Gibson2013}. The results presented here add confidence to the fact that stars and nebulae can provide valid and coherent constraints to the chemical evolution of the Galaxy and that we understand the differences in their abundances. The results presented in this paper lead the way to future usage of nebular and stellar samples together to study the chemical evolution of the Milky Way, and, since the Galaxy can be a benchmark for other star-forming galaxies in the context of the chemical evolution of galaxies.

In the future, 
we will use the Galactic RGs and PN abundances to study and compare their metallicity gradients, in the framework of other spiral galaxies \citep[e.~g., in M31,][]{2022MNRAS.517.2343B}. 
We will also extend the study of this paper to PNe and RGs in the Small and Large Magellanic Clouds, where we will be able to test the strength of the connection between nebular and stellar abundances at varying galaxy metallicity, which is extremely important when studying external galaxies.

\appendix \section{}

\begin{figure}[htb!]
   \centering
   \includegraphics[width=18cm]{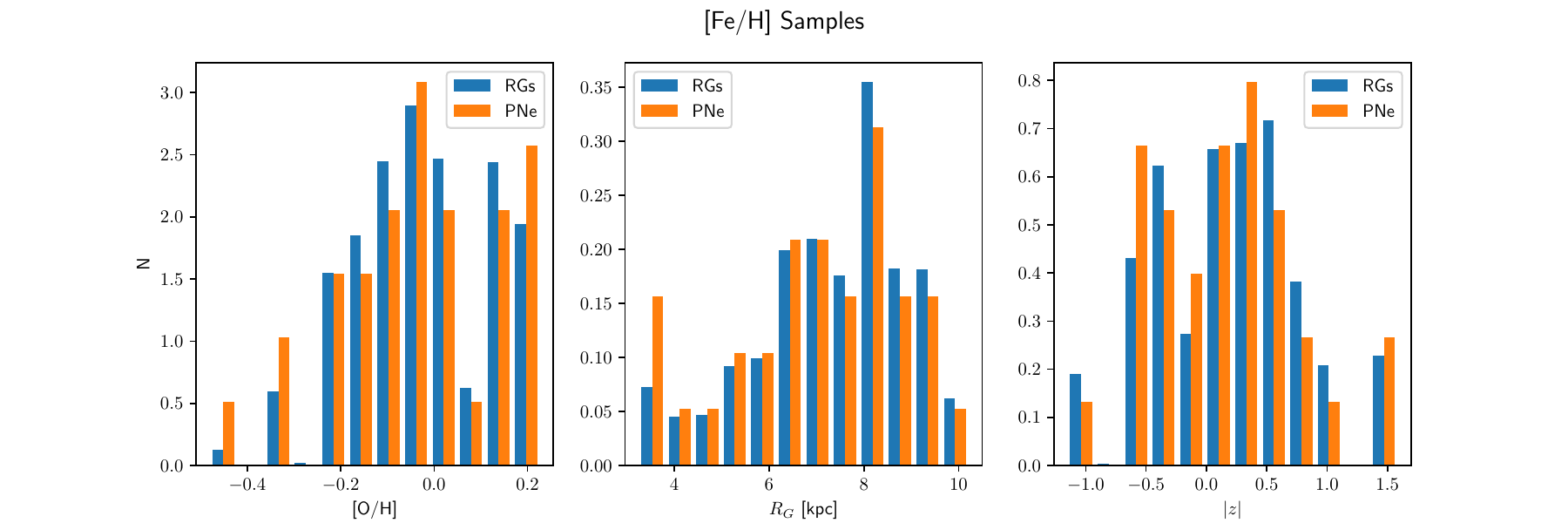}
      \caption{As in Fig.~\ref{3distr}, but for the [Fe/H] Samples.}
      \label{3distr_fe}
    \end{figure}

\begin{figure}[htb!]
   \centering
   \includegraphics[width=18cm]{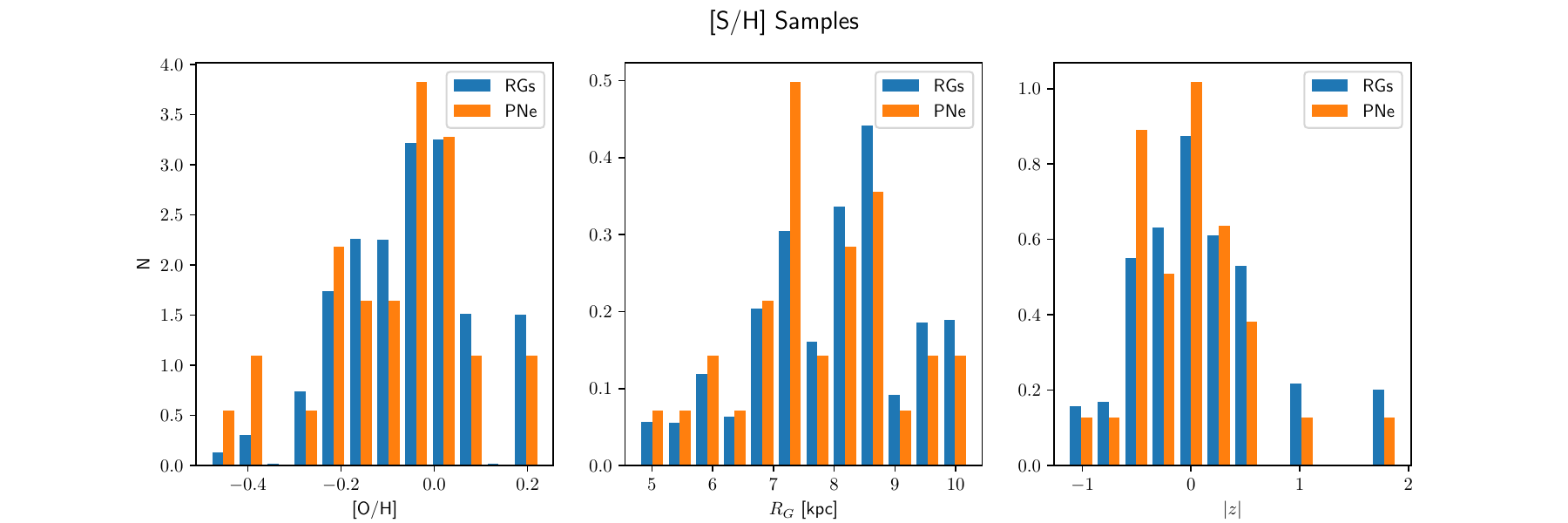}
      \caption{As in Fig.~\ref{3distr}, but for the [S/H] Samples.}
      \label{3distr_sul}
    \end{figure}

\begin{figure}[htb!]
   \centering
   \includegraphics[width=18cm]{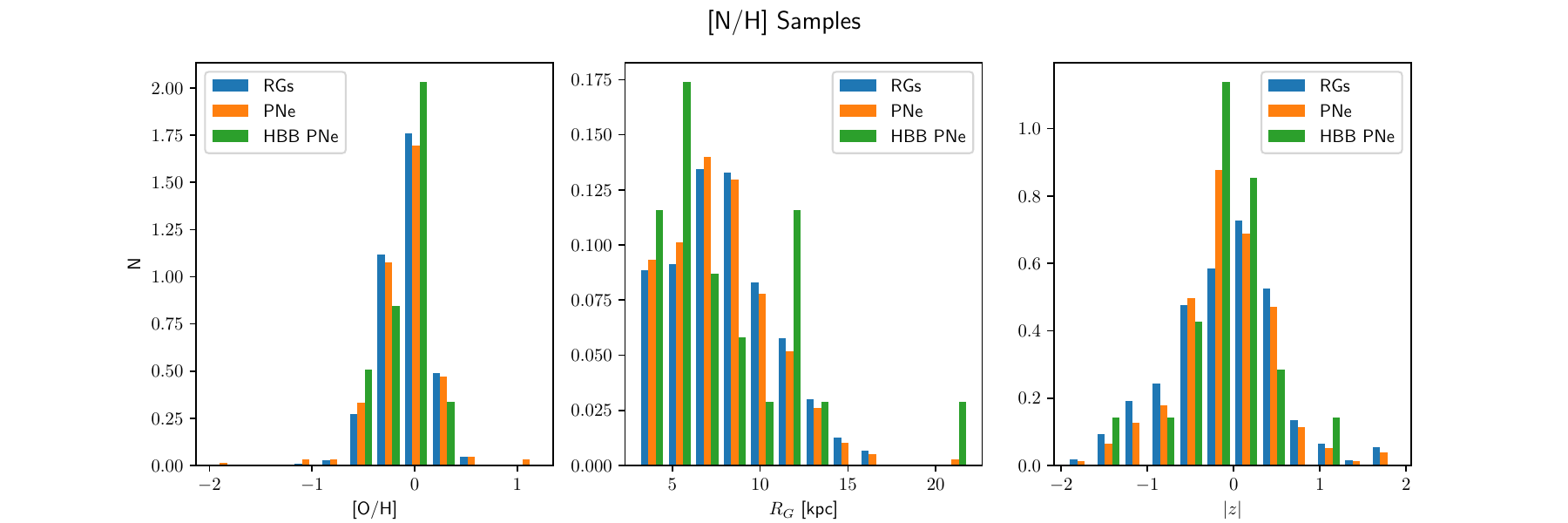}
      \caption{As in Fig.~\ref{3distr}, but for the [N/H] RG and PN samples, and the HBB PN sample.}
      \label{3distr_nit}
    \end{figure}

\begin{figure}[htb!]
   \centering
   \includegraphics[width=18cm]{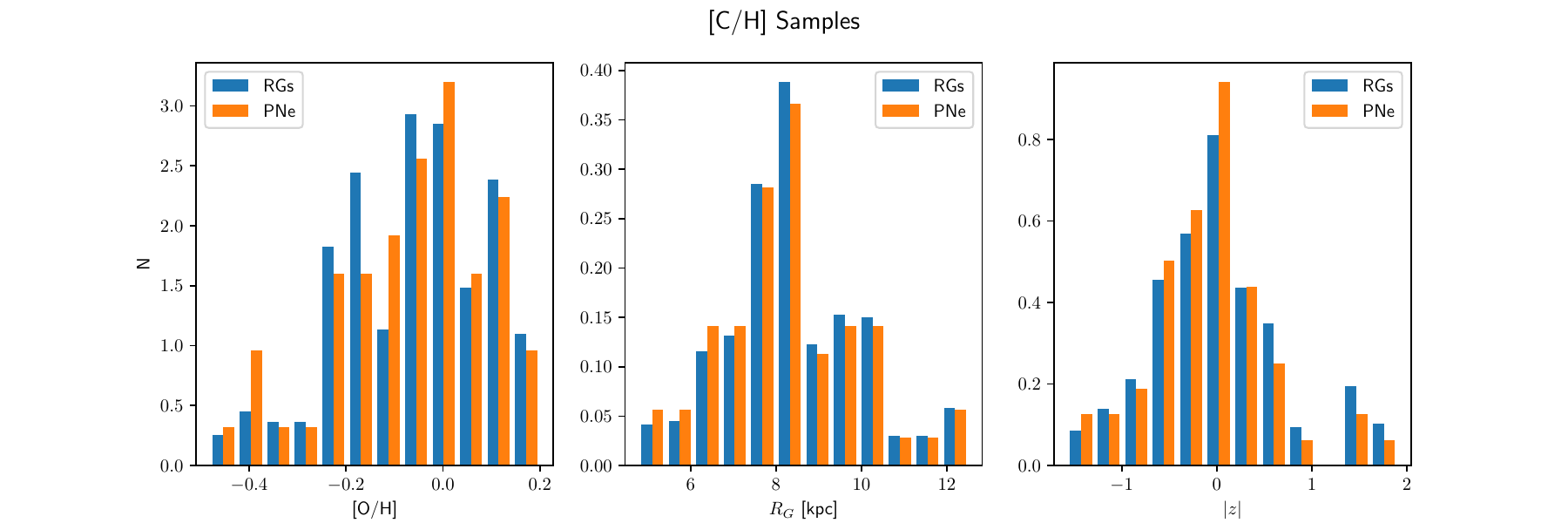}
      \caption{As in Fig.~\ref{3distr}, but for the [C/H] Samples.}
      \label{3distr_car}
    \end{figure}

Figures~\ref{3distr_fe}, \ref{3distr_sul}, \ref{3distr_nit}, and \ref{3distr_car} show the parameter distributions as in Fig.\ref{3distr} but for the [Fe/H], [S/H], [N/H], and [C/H] Samples respectively. 

\section{}
In \citet{KB94}'s sulfur abundance determinations, they derive S$^{2+}$ from the S$^+$ (their Eq. A38) when [\ion{S}{3}] was not observed. Several other data sets use this approximation directly from \citet{KB94}, whose ICF scheme is commonly used in the literature.
Eq. A38 in \citet{KB94} is based on their Fig. 15. But, in fact, Eq.~A38 is incorrect, and it should instead read as: 
${\rm S^{2+}/S^+=4.677\times(O^{2+}/O^+)^{0.433}}$. As written in \citet{KB94}, it underestimates sulfur abundances. The correction factor to be added to the published abundances that use this method is:
${\rm log((1+4.677\times(O^{2+}/O^+)^{0.433})/(1+4.677+(O^{2+}/O^+)^{0.433}))}$, which is between 0 and 0.38 for typical values of O$^{2+}$/O$^+$.

We modify the sulfur abundance selection from the references of Table~\ref{table:PN_abref} to allow for the newly found issue with the ICF. The data set by \citet{PBS10} does not use ICFs, since they observe the [\ion{S}{3}] IR line. The same is true for most of the \citet{TBL03} and \citet{MHB19} sulfur abundances, where the [\ion{S}{3}] lines are in all but a couple of PNe that we excluded from the following analysis. We also excluded all sulfur abundances from \citet{GHG14} since they used the ICF from \citet{KB94}, but we could not find the information on which PNe these were used. Finally, from \citet{KB94}, we used only the sulfur abundances for PNe whose [\ion{S}{3}] line is available. 

\section{Acknowledgements}
This paper uses public data from the SDSS IV APOGEE survey. Funding for the Sloan Digital Sky Survey IV has been provided by the Alfred P. Sloan Foundation, the U.S. Department of Energy Office of Science, and the Participating Institutions. SDSS-IV acknowledges support and resources from the Center for High-Performance Computing at the University of Utah. The SDSS web site is www.sdss.org.
SDSS-IV is managed by the Astrophysical Research consortium for the Participating Institutions of the SDSS Collaboration including the Brazilian Participation Group, the Carnegie Institution for Science, Carnegie Mellon University, the Chilean Participation Group, the French Participation Group, Harvard-Smithsonian Center for Astrophysics, Instituto de Astrof\'isica de Canarias, The Johns Hopkins University, 
Kavli Institute for the Physics and Mathematics of the Universe (IPMU) /  University of Tokyo, Lawrence Berkeley National Laboratory, Leibniz Institut f\"ur Astrophysik Potsdam (AIP),  Max-Planck-Institut f\"ur Astronomie (MPIA Heidelberg), Max-Planck-Institut f\"ur Astrophysik (MPA Garching), Max-Planck-Institut f\"ur Extraterrestrische Physik (MPE), National Astronomical Observatory of China, New Mexico State University, New York University, University of Notre Dame, Observat\'orio Nacional / MCTI, The Ohio State University, Pennsylvania State University, Shanghai Astronomical Observatory, United Kingdom Participation Group,
Universidad Nacional Aut\'onoma de M\'exico, University of Arizona, University of Colorado Boulder, University of Oxford, University of Portsmouth, University of Utah, University of Virginia, University of Washington, University of Wisconsin, Vanderbilt University, and Yale University.

\end{document}